\documentclass[aps,prl,twocolumn,superscriptaddress,showpacs,floatfix,twocolumn]{revtex4-1}

\usepackage[pdftex]{graphicx,graphics}

\usepackage{mathrsfs}
\usepackage{amsmath,amsfonts,amsbsy,amssymb,amsthm}
\usepackage{mathrsfs,bbm}
\usepackage{epstopdf}
\usepackage[colorlinks=true,
            urlcolor=blue,
            citecolor=red,
            pdfstartview=FitH,
            pdfpagemode=None]{hyperref}

\makeatletter
\@ifundefined{textcolor}{}
{%
 \definecolor{BLACK}{gray}{0}
 \definecolor{WHITE}{gray}{1}
 \definecolor{RED}{rgb}{1,0,0}
 \definecolor{GREEN}{rgb}{0,1,0}
 \definecolor{BLUE}{rgb}{0,0,1}
 \definecolor{CYAN}{cmyk}{1,0,0,0}
 \definecolor{MAGENTA}{cmyk}{0,1,0,0}
 \definecolor{YELLOW}{cmyk}{0,0,1,0}
}

\newcommand{\beq}{\begin{equation}}
\newcommand{\eneq}{\end{equation}}
\newcommand{\beqnn}{\begin{equation*}}
\newcommand{\eneqnn}{\end{equation*}}
\newcommand{\beqy}{\begin{eqnarray}}
\newcommand{\eneqy}{\end{eqnarray}}
\newcommand{\beqynn}{\begin{eqnarray*}}
\newcommand{\eneqynn}{\end{eqnarray*}}

\newcommand{\expect}[1]{\big\langle #1 \big\rangle}
\newcommand{\erf}[1]{Eq. (\ref{#1})}

\begin{document}

\title[]
{Qubit noise spectroscopy for non-Gaussian dephasing environments} 

\author{Leigh M. Norris}
\affiliation{ \mbox{Department of Physics and Astronomy, Dartmouth College,
6127 Wilder Laboratory, Hanover, New Hampshire 03755, USA}}

\author{Gerardo Paz-Silva}
\affiliation{{ Centre for Quantum Dynamics \& 
Centre for Quantum Computation and Communication Technology,  
Griffith University, Brisbane, Queensland 4111, Australia}} 

\author{Lorenza Viola}
\affiliation{ \mbox{Department of Physics and Astronomy, Dartmouth College,
6127 Wilder Laboratory, Hanover, New Hampshire 03755, USA}}

\date{\today}

\begin{abstract}
We introduce open-loop quantum control protocols for characterizing the spectral properties of non-Gaussian 
noise, applicable to both classical and quantum dephasing environments. The basic idea is to engineer a 
multi-dimensional frequency comb via repetition of suitably designed pulse sequences, through which the desired 
high-order noise spectra may be related to observable properties of the qubit probe.   We prove that 
access to a high time resolution is key to achieve spectral reconstruction over an extended bandwidth, 
overcoming limitations of existing schemes.  Non-Gaussian spectroscopy is demonstrated for a classical noise 
model describing quadratic dephasing at an optimal point, as well as a quantum spin-boson model out of equilibrium. 
In both cases, we obtain spectral reconstructions that accurately predict the qubit dynamics in the non-Gaussian regime.
\end{abstract}


\maketitle

Accurately characterizing the spectral properties of environmental noise in open quantum systems 
has broad practical and fundamental significance.
Within quantum information processing, 
this is a prerequisite for optimally tailoring the design of quantum control and error-correcting strategies to the 
noisy environment that qubits experience, and for testing key 
assumptions in fault-tolerance threshold  
derivations \cite{PreskillSufficient}.
From a physical standpoint, precise knowledge of the noise
is necessary for quantitatively modeling and understanding open-system dynamics, 
with implications ranging from the classical-to-quantum 
transition to non-equilibrium quantum statistical mechanics and quantum-limited metrology \cite{Breuer:book}.   

Quantum noise spectroscopy seeks to characterize the spectral properties of environmental noise by using a 
controlled quantum system (a qubit under multi-pulse control in the simplest case) as a dynamical probe \cite{OlderRefs}.  
In recent years, interest in quantum noise spectroscopy has heightened thanks to both improved theoretical understanding 
of open-loop controlled dynamics in terms of transfer filter-function (FF) techniques \cite{MikeFF,Paz2014} and experimental 
validation in different qubit platforms.  
In particular, quantum control protocols based on dynamical decoupling (DD)  
have been successfully implemented to characterize noise properties during memory and 
driven evolution in systems as diverse as solid-state nuclear magnetic resonance 
\cite{Alvarez2011}, superconducting \cite{Oliver} and spin \cite{SpinQubits} qubits, and  
nitrogen vacancy centers in diamond \cite{NVs}.

Despite the above advances, existing noise spectroscopy protocols suffer from several disadvantages.  
Notably, they are restricted to {\em classical, Gaussian} phase noise. While dephasing ($T_2$-) processes are 
known to provide the dominant decoherence mechanism in a variety of realistic scenarios, the Gaussianity assumption 
is {\em a priori} far less justified.  On the one hand, the Gaussian approximation tends to break down in 
situations where the system is strongly coupled with an environment consisting of discrete degrees of freedom --  
such as for $1/f$ noise, as ubiquitously encountered in solid-state devices \cite{Paladino}. 
Even for environments well described by a continuum of modes, 
non-Gaussian noise statistics may be generally expected away from thermal equilibrium, or whenever symmetry 
considerations forbid a linear coupling 
\cite{quadratic}. In all such cases, accurate noise spectroscopy mandates going beyond the Gaussian regime.

In this paper, we introduce open-loop control protocols for characterizing {\em stationary, non-Gaussian dephasing} 
using a qubit probe.  Our approach is applicable to classical noise environments and
to a paradigmatic class of open quantum systems described by linearly coupled oscillator environments -- as long as  
all relevant noise spectra obey suitable smoothness assumptions. 
While we build on the noise spectroscopy by sequence repetition proposed by Alvarez and Suter \cite{Alvarez2011},
our central insight is to leverage the simple structure of FFs in a purely dephasing setting to establish the emergence of a
frequency comb for arbitrary high-order noise spectra (so-called $polyspectra$), paving the way to the desired 
multi-dimensional spectral estimation.
We first demonstrate the power of our approach for Gaussian noise, where we extend the range of spectral reconstruction 
over existing protocols. In the non-Gaussian regime, we reconstruct the spectra associated with the leading high-order cumulants of the noise, absent in the Gaussian limit.  Quantitative prediction of the qubit free evolution in the presence of these non-Gaussian environments reveals clear dynamical signatures in both the classical and quantum case. 

{\em Control setting and noise polyspectra.--} We consider a qubit $S$ coupled to an uncontrollable environment (bath) $B$. In the interaction picture with respect to the bath Hamiltonian, $H_B$, and the qubit Hamiltonian  $H_S=\hbar\omega_0 \sigma_z/2$, the joint system is described by $H(t)=\hbar\sigma_z B(t)/2+H_{\text{ctrl}}(t)$, where
the first term accounts for the bath-induced dephasing and $H_{\text{ctrl}}(t)$ is the external open-loop control, 
acting non-trivially on the qubit alone. For a classical bath, $B(t)$ is a stochastic noise process, whereas 
$B(t)$ is a time-dependent operator for a quantum bath.
The applied control consists of repeated sequences of $\pi$-pulses (say, about $x$), which for simplicity we take to be 
instantaneous. After transforming to the interaction picture associated with $H_{\text{ctrl}}(t)$, 
the joint Hamiltonian becomes $\tilde{H}(t)= y(t) \hbar \sigma_zB(t)/2$, where 
the ``switching function" $y(t)$ changes sign between $\pm 1$ with every $\pi$-pulse applied to the qubit. 

The effect of dephasing is seen in the dynamics of the qubit's coherence element, 
which we may express in terms of bath-operator cumulants. Specifically, 
$\langle \sigma_+ (t) \rangle = \langle \sigma_+ (0) \rangle \,\text{e}^{{-\chi(t)} +i \phi(t)}$, 
where the decay parameter and phase angle are respectively given by:
\begin{align}
\chi(t) =\sum_{\ell=1}^\infty\!\frac{(-1)^\ell}{(2\ell)!}\Upsilon^{(2\ell)}(t), \; 
\phi(t)=\sum_{\ell=1}^\infty\!\frac{(-1)^\ell}{(2\ell+1)!}\Upsilon^{(2\ell+1)}(t), \nonumber \\
\hspace*{-4mm}
\Upsilon^{(k)}(t) \equiv \int_0^{t}\!\! dt_1\ldots \!\!\int_0^{t}\!\!dt_{k}\,
y(t_1)\ldots y(t_{k})C^{(k)}(t_1,\ldots ,t_{k}) , \hspace*{3mm}
\nonumber
\end{align}
where the $k$th-order noise cumulant $C^{(k)}(t_1,\ldots ,t_k)$ depends on 
the bath correlation functions $\expect{B(t_1) \ldots B(t_j)}$, $j\leq k$, and $\expect{\cdot}$ denotes 
an ensemble average for a  classical bath or an expectation value with respect to the initial bath state, 
$\rho_B(0)$, in the quantum case.  
For zero-mean Gaussian noise, $C^{(k)}(t_1,\ldots,t_{k}) \equiv 0$ except for $k= 2$. Thus, Gaussian noise 
gives {\em no phase evolution}. For non-Gaussian noise, 
higher-order even (odd) cumulants contribute to decay (phase evolution), respectively. 

For stationary noise,  where $C^{(n+1)}(t_1,\ldots,t_{n+1})$ is a function of the time separations 
$\tau_j \equiv t_{j+1}-t_1$, $j\in \{ 1,\ldots, n\}$, the noise spectral properties are fully characterized by the 
Fourier transforms of the cumulants with respect to $\{ \tau_j \}$. 
Using the compact notation $\vec{v}_n \equiv (v_1,\ldots,v_n)$, 
the {\em $n$th-order $polyspectrum$} is defined as
\begin{align}
&S_{n}(\vec{\omega}_{n}) \equiv \!\!\!\int_{\mathbb{R}^n}\!\!\!d\vec{\tau}_{n} \,
e^{-i \vec{\omega}_{n}\cdot\vec{\tau}_{n}} C^{(n+1)}(\vec{\tau}_{n}), \quad n \geq 1, 
\label{eq::polyspectra}
\end{align}
where $S_1(\vec{\omega}_1)\equiv S(\omega)$ is the familiar power spectral density (PSD), and 
$S_2(\vec{\omega}_2)$, $S_3(\vec{\omega}_3)$ are known as the ``bi-spectrum'' and ``tri-spectrum'', respectively. 
For all orders, $S_{n}(\vec{\omega}_{n}\!)$ is a smooth $n$-dimensional surface when the noise is classical and 
ergodic \cite{Brillinger}. 
More generally, $C^{(n+1)}(t_1,\ldots t_{n+1})$ may depend on fewer than $n$ time separations, leading to the presence
of delta functions in $S_{n}(\vec{\omega}_{n})$. All polyspectra possess a high degree of symmetry, {\em irrespective of the noise}.  
That is, $S_{n}(\vec{\omega}_{n})$ is fully specified in all frequency space by its value on a particular subspace, 
${\mathcal D}_n$, known as the \emph{principal domain} \cite{Chandran1994}.  

{\em Noise spectroscopy protocol.--} 
Our objective is to characterize not only the PSD but the polyspectra. 
We accomplish this by adapting the DD noise spectroscopy protocol proposed in \cite{Alvarez2011} for Gaussian noise.  
This protocol relies on repetitions of identical {\em base sequences}, whose duration (``cycle time'') we shall denote by $T$. 
Following \cite{Paz2014}, the effect of a base control sequence $p$ in the frequency domain is characterized by a
single {\em fundamental FF}, $F_p(\omega) \equiv \int_0^Tdte^{ + i\omega t} y_p(t)$.
If $|\vec{\omega}_{k-\!1}| \equiv \omega_1+ \ldots +\omega_{k-\!1}$, direct calculation shows that $M$ repetitions of 
$p$ yield
\begin{align}
\Upsilon_{[p]^M}^{(k)}
\!=&\!\!\!\int_{\mathbb{R}^{k-\!1}}\!\!\!\!\!\!\!d\vec{\omega}_{k-\!1}\!\!
\prod_{j=1}^{k-\!1}\!\!F_p(\omega_j)\frac{\text{sin}(M\omega_jT/2)}{\text{sin}(\omega_jT/2)}\notag
\\&\times F_p(-|\vec{\omega}_{k-\!1}|)\frac{\text{sin}(M|\vec{\omega}_{k-\!1}|T/2)}{\text{sin}(|\vec{\omega}_{k-\!1}|T/2)}\frac{S_{k-\!1}(\vec{\omega}_{k-\!1})}{(2\pi)^{k-\!1}},
\label{eq::FTcumulantM}
\end{align}
The key to extending the protocol in \cite{Alvarez2011}
beyond Gaussian noise ($k=2$) is to realize that {\em repetition produces a multi-dimensional frequency comb for all orders}, namely, 
\begin{align}
\prod_{j=1}^{k-1}\!&\Big[\frac{\text{sin}(M\omega_jT/2)}{\text{sin}(\omega_jT/2)}\Big]\frac{\text{sin}(M|\vec{\omega}_{k-1}|T/2)}{\text{sin}(|\vec{\omega}_{k-1}|T/2)}
\label{eq::hypercomb}
\\&\approx\!M\prod_{j=1}^{k-1}\!\!\Big[\frac{2\pi}{T}\!\!\sum_{n_j\!=-\!\infty}^{\infty}\!\!\!\delta\Big(\omega_{j}\!-\!\frac{2\pi n_j}{T}\Big)\Big]\notag , \quad M \gg1, \forall k, 
\end{align}
provided that $S_{k-\!1}(\vec{\omega}_{k-\!1})$ in Eq. (\ref{eq::FTcumulantM}) is a smooth function.

Thanks to the ``hyper-comb''  in \erf{eq::hypercomb}, obtaining the polyspectra becomes an inverse problem. Substituting \erf{eq::hypercomb} into \erf{eq::FTcumulantM} produces a linear equation that couples the polyspectra and the FFs evaluated at the harmonic frequencies 
${\cal H}_j \equiv \{2\pi\vec{n}_j/T\,|\,\vec{n}_j\in\mathbb{Z}^j\}$,
\begin{align}
\Upsilon_{[p]^M}^{(k)}\!\!=\!\!\!\!\!\!\!\!\!\sum_{\vec{h}_{k-\!1}\in {\cal H}_{k-\!1}}\!\!\!\!\!\!\!\!\frac{M}{T^{k-\!1}}\!\!\prod_{j=1}^{k-1}\!\!F_p(h_j)
F_p(-|\vec{h}_{k-\! 1}|)S_{k-\!1}(\vec{h}_{k-\!1})
\label{eq::FTlinear}.
\end{align}
To obtain a finite linear equation, we need to truncate the above sum to a finite set $\Omega_{k-\!1}$. 
With no prior knowledge of the noise, it suffices to consider
$\Omega_{k-\!1} \subset {\cal D}_{k-1} \cap {\mathcal H}_{k-\!1}$ 
in the principal domain of the polyspectrum. 
Truncating the expression in \erf{eq::FTlinear} 
enables us to relate the sampled polyspectra 
to experimentally observable dynamical quantities:
\begin{align}
\chi_{[p]^M}   
\approx&\sum_{\ell=1}^\infty\frac{(-1)^\ell M}{(2\ell)!\,T^{2\ell-\!1}}\hspace*{-3mm}
\sum_{\vec{h}_{2\ell-\!1}\in\Omega_{2\ell-\!1}}\!\!\!\!\!m_{2\ell-\!1}(\vec{h}_{2\ell-\!1}) \nonumber \\
&\times\prod_{j=1}^{2\ell-\!1}\!\!F_p(h_j)
F_p(-|\vec{h}_{2\ell-\!1}|)S_{2\ell-1}(\vec{h}_{2\ell-\!1}) , 
\label{eq::chiM} \\
\phi_{[p]^M} 
\approx&\sum_{\ell=1}^\infty\frac{(-1)^\ell M}{(2\ell+1)!\,T^{2\ell}}\sum_{\vec{h}_{2\ell}\in\Omega_{2\ell}}m_{2\ell}(\vec{h}_{2\ell})
\nonumber \\
&\times\prod_{j=1}^{2\ell}F_p(h_j)
F_p(-|\vec{h}_{2\ell}|)S_{2\ell}(\vec{h}_{2\ell}),
\label{eq::phiM}
\end{align}
where the multiplicity $m_n(\vec{h}_n) \equiv \text{card}\{ \vec{h}_n\in {\mathbb R}^n \,|\, S_n(\vec{h}_n)\\ = 
S_n(\vec{\omega}_n), \, \forall 
\omega_n \in {\mathcal D}_n\}$ accounts for the symmetry of the polyspectrum.
Whenever the contributions from high-order multi-point correlation functions are negligible 
(e.g., for sufficiently small evolution time 
and/or noise strength), the cumulant expansion in Eqs. (\ref{eq::chiM})-(\ref{eq::phiM}) 
may be truncated at a finite $\ell=L$. If $N$ terms remain after truncation, experimentally measuring 
$\chi_{[p]^M}$ ($\phi_{[p]^M}$) for at least $N$ control sequences 
creates a system of linear equations, that can be inverted to obtain the odd (even) polyspectra up to order $2L-1$ ($2L$)
\cite{Note1}.

{\em Base sequence construction.--} 
In the original noise spectroscopy protocol of \cite{Alvarez2011}, a {\em fixed} base sequence is used 
(CPMG, after Carr, Purcell, Meiboom, and Gill), 
with cycle times varying from $T$ to $T/n=2\tau$, where $n\in\mathbb{Z}^{+}$ and $\tau$ is the minimum time between pulses. 
While this produces a well-conditioned linear inversion, both the number of distinct control sequences 
and the range of spectral reconstruction are limited -- in particular, 
$|\omega| \leq \pi/\tau$ for a minimum allowed $\tau >0$. 
The use of a fixed DD sequence has an additional disadvantage: CPMG 
refocuses static noise ($F_{\text{cpmg}}(\omega=0)=0$, hence the 
``filtering order'' is non-zero \cite{Paz2014}), precluding reconstruction at any point in 
frequency space containing a zero, a substantial information loss for higher-dimensional polyspectra.  

Non-Gaussian noise spectroscopy demands a {\em large number of sequences with spectrally distinct FFs}, 
including some with zero filtering order.  We generate a family of base sequences satisfying these requirements by 
using different orders of concatenated DD, CDD$_m$: namely, not only CPMG ($m=2$), but also durations of 
free evolution ($m=0$), up to maximum DD order $m=5$.  The presence of free evolution permits sequences 
with zero filtering order, enabling the polyspectra to be reconstructed at points containing a zero. 
Specifically, let a \emph{fixed} cycle time $T$ be expressed in terms of a minimum {\em time resolution} $\delta$,  
$T \equiv q \delta$, where $q \in\mathbb{Z}^{+}$.  While all pulse times will be integer multiples of $\delta$, 
$\delta$ and $\tau$ are two independent constraints a priori, with $\delta <\tau$ in typical settings.
If $q \equiv \sum_i q_i$ is an integer partition of $q$, we place a CDD$_m$ sequence into the $i$th interval, 
of duration $q_i \delta$, subject to the condition that no two pulses are separated by less than $\tau$.  
As shown in the Supplement \cite{Supplement}, the range of spectral reconstruction is bounded by $|\omega|\leq\pi/\delta$. 
A high resolution (small $\delta$) is key to generate sequences with {\em incommensurate} periodicities, making it possible 
to achieve spectral reconstruction over an extended range.  

\begin{figure}[t]
\includegraphics[width=0.83\columnwidth]{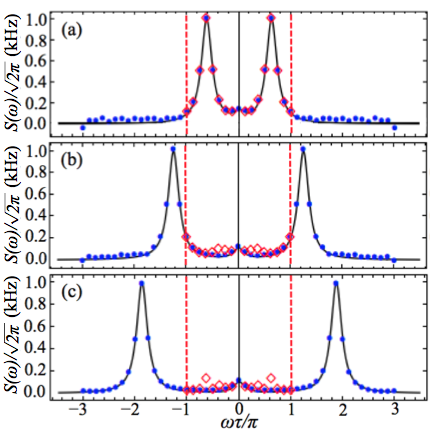}  
\vspace*{-2mm}
\caption{
(Color online) Comparison between Alvarez-Suter DD spectroscopy protocol (red diamonds) and the proposed 
protocol (blue dots) in reconstructing a Gaussian noise PSD with increasing high-frequency components (top to bottom).
Both protocols use $M=50$ repetitions of control sequences 
with $\tau=3.1\times 10^{-4}\,$s and $T=16\tau$. For our protocol, 
we employ 25 base sequences assembled from CDD$_m$, $m=0, \ldots, 4$.
The spectrum is a sum of Lorentzians peaks, 
$S(\omega) ={w_1}/{[1+ (8\omega/\omega_c)]^2} + {w_2}/\{1 + [8\,(\text{sign}(\omega)\omega-d)/\omega_c)^2]\}$, 
where $w_1/\sqrt{2\pi}=0.1\,\text{kHz}$, $w_2/\sqrt{2\pi}=1\,\text{kHz}$, $\omega_c=10\,\text{kHz}\approx{\pi}/{\tau}$, 
and $d$ controls the offset of the high-frequency peaks, 
$d=\frac{5}{8}{\pi}/{\tau}$ (a), $d=\frac{10}{8} {\pi}/{\tau}$ (b), 
$d=\frac{15}{8} {\pi}/{\tau}$ (c). 
As the original protocol can only reconstruct $S(\omega)$ up to $|\omega|<{\pi}/{\tau}$ (dashed vertical lines), 
it cannot ``see" the high-frequency peaks in (b)-(c), which results in instability at lower frequencies. 
}
\label{fig::Lorentzian3PiTau}
\end{figure}

The added capabilities of our control sequences may be appreciated already for spectroscopy of classical Gaussian 
noise, see Fig. \ref{fig::Lorentzian3PiTau}. 
In this case, \erf{eq::chiM} truncates exactly at $\ell=1$; this produces a system of linear equations relating the 
desired PSD to $\chi_{[p]^M}$, which we obtain numerically for each control sequence. 
In addition to accurately reconstructing the larger peaks over the expanded range $|\omega|\leq 48\pi/T=3\pi/\tau$, our protocol 
successfully resolves the small peak at $\omega=0$, thanks to inclusion of control sequences with zero filtering order. 

\begin{figure*}[ht]
\includegraphics[scale=.5]{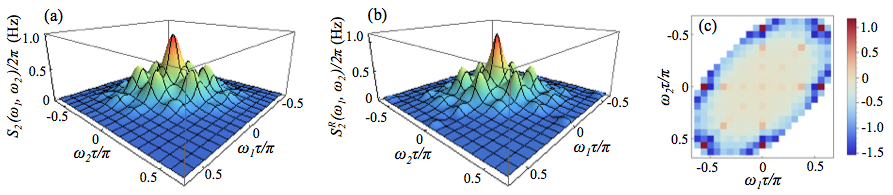}
\vspace*{-3mm}
\caption{
(Color online) Actual bispectrum, $S_2(\vec{\omega}_2)$ (a), vs. reconstructed bispectrum, 
$S_2^R( \vec{\omega}_2)$ (b), and relative error 
$E(\vec{\omega}_2)\equiv [S_2^R( \vec{\omega}_2)-S_2(\vec{\omega}_2)]/S_2(\vec{\omega}_2)$ (c), 
for classical non-Gaussian square noise $\xi_1 (t) = g(t)^2-\expect{g(t)^2}$.
Here, $g(t)$ is Gaussian with spectrum 
$S_g(\omega)={w_1}/[{1+8(\omega/\omega_c)^2}]+{w_2}/{[1+16\, (\text{sign}(\omega)\omega/\omega_c-3/2)^2]}$, 
and $w_1=1/10$ Hz, $w_2=1/25$ Hz. The  protocol uses 
$M=40$ repetitions of sequences composed of CDD$_{0-5}$, with  $\tau=3.95\times 10^{-5}$ s and 
$T=32\tau$ to reconstruct the bi-spectrum at 325 points.
In (b), these values have been smoothed with a spline interpolation.  
The largest relative errors occur in the high-frequency regions at the outer edge of the bi-spectrum which, 
however, contribute far less to the qubit dynamics. 
\label{fig:classicalBispectrum}
}
\end{figure*}

\begin{figure}[h]
\center
\includegraphics[width=1\columnwidth]{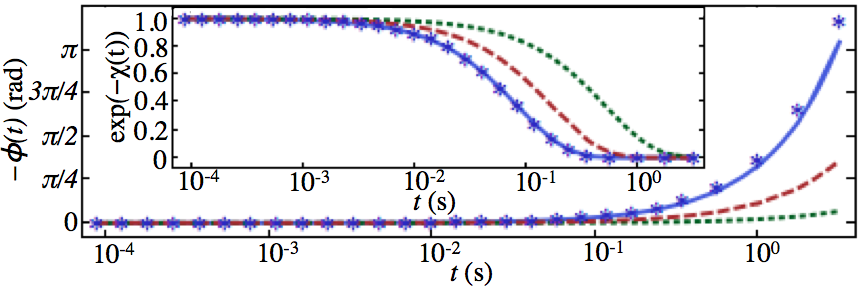}
\vspace*{-5mm}
\caption{
(Color online) Phase evolution and decay (inset) of a qubit under square noise $\xi_a(t)$ 
with different degrees of Gaussianity [see text], 
and same spectrum $S_g(\omega)$ for $g(t)$ as in Fig. \ref{fig:classicalBispectrum}.
Curves are ordered according to decreasing degree of non-Gaussianity: $a=1$ (blue solid), $a=0.7$ 
(red dashes) and $a=0.4$ (green dots). For the fully non-Gaussian 
$a=1$ case, we have used the reconstructed spectrum and 
bi-spectrum [Fig. \ref{fig:classicalBispectrum}] to predict the qubit decay and phase evolution (blue asterisks), showing 
excellent agreement with the theoretical evolution, computed up to the fifth-order noise cumulant. }
\label{fig::FID}
\end{figure}

{\em Non-Gaussian spectral reconstructions.--}
We now return to our main goal, namely characterizing non-Gaussian polyspectra. As a first example, we consider 
a classical ``square noise" process 
arising from a quadratic coupling to a Gaussian source, as encountered in superconducting qubits operating at an 
optimal working point \cite{Paladino,CywinskiOptimal}.  That is, 
$B(t)\equiv\xi_a (t)=a\big(g(t)^2-\expect{g(t)^2}\big)+\big(1-a\big)g(t)$,
where $g(t)$ is a zero-mean Gaussian process, 
and $a\in[0,1]$ interpolates between Gaussian ($a=0$) and fully non-Gaussian ($a=1$) regimes. 
Truncating \erf{eq::phiM} at the leading $\ell=1$ term allows us to reconstruct the bi-spectrum $S_2(\vec{\omega}_2)$ 
from numerically determined values of $\phi_{[p]^M}$. Here, 
the relevant principal domain ${\mathcal D}_2$ is an octant bounded by 
$\omega_1=\omega_2$ and $\omega_1=0$. 
Reconstructing 35 points in ${\cal D}_2$ enables us to obtain $S_2(\vec{\omega}_2)$ at 325 points in $\mathbb{R}^2$. 
Representative results for the actual vs. reconstructed bi-spectrum at $a=1$ are shown in Fig. 
\ref{fig:classicalBispectrum}(a)-(b). The relative error in \ref{fig:classicalBispectrum}(c) indicates very good agreement at the interior points,  
but larger error in the tails. Because there is minimal spectral concentration in the tails, however, this error has little  effect on the qubit dynamics.  As Fig. \ref{fig::FID} shows, excellent agreement is found between the 
theoretical phase evolution and the one predicted by the reconstructed bi-spectrum. 

Extending quantum noise spectroscopy methods to quantum environments entails qualitatively new challenges because 
non-Gaussian statistics ensues now from the {\em combined effect of the bath operators $B(t)$ and the initial bath state} 
$\rho_B(0)$, and no general characterization of quantum polyspectra (and their smoothness properties) 
is available to the best of our knowledge.  
We take a first step in this direction by focusing on 
linearly coupled spin-boson environments, in which case $H_B=\hbar\sum_k \Omega_k a_k^\dag a_k$ 
and $B(t)=\sum_k (g_k e^{i\Omega_k t} a_k^\dag +\text{h.c.})$, where $a_k, a_k^\dag$ are 
canonical bosonic operators and both $\Omega_k, g_k$ have units of (angular) frequency. 
For a general quantum bath, the noise is stationary if and only if $[H_B, \rho_B(0) ]=0$.  This prevents non-zero 
odd cumulants in the spin-boson model, implying that the qubit undergoes no phase evolution. 

Given any stationary, non-Gaussian bath state $\rho_B(0)$ we can, 
using our protocol, reconstruct spectral quantities associated with the first two leading-order even cumulants, $S(\omega)$ and 
$S_3(\vec{\omega}_3)$.
Although $S(\omega)$  is {\em asymmetric} about $\omega=0$, the fact that arbitrary FFs enter through 
even combinations implies that we can only reconstruct its symmetric component, or ``effective spectrum",  
$S_{\text{eff}}(\omega)  \equiv [S(\omega)+S(-\omega)]/2$. $S_{\text{eff}}(\omega)$ is the quantity relevant to the qubit dynamics. 
As shown in \cite{Supplement}, the tri-spectrum for any {\em non-separable}, stationary initial bath state has the form 
$S_3 (\vec{\omega}_3)\!=(2\pi)^3[\delta(\omega_1+\omega_2)J_3(\omega_1,\omega_3)+\delta(\omega_2+\omega_3)J_3(\omega_1,\omega_2)
+\delta(\omega_1\!+\!\omega_3)J_3(\omega_2,\omega_3)].$
Because the hypercomb approximation holds only if $S_3(\vec{\omega}_3)$ is smooth, we cannot directly reconstruct it. 
We can, however, reconstruct the ``effective tri-spectrum" $J_3(\vec{\omega}_2)$, provided it is smooth. 
Due to the delta functions in $S_3(\vec{\omega}_3)$, the terms in \erf{eq::chiM} associated with the tri-spectrum differ 
by a constant factor in the spin-boson case. The modified equations are derived in \cite{Supplement}, along with similar 
equations for {\em separable} stationary initial states.
In the absence of prior information about $\rho_B(0)$, comparison between predictions based on the two 
resulting reconstructions will enable the correct effective tri-spectrum to be inferred. 

\begin{figure}[t]
\center
\includegraphics[width=1\columnwidth]{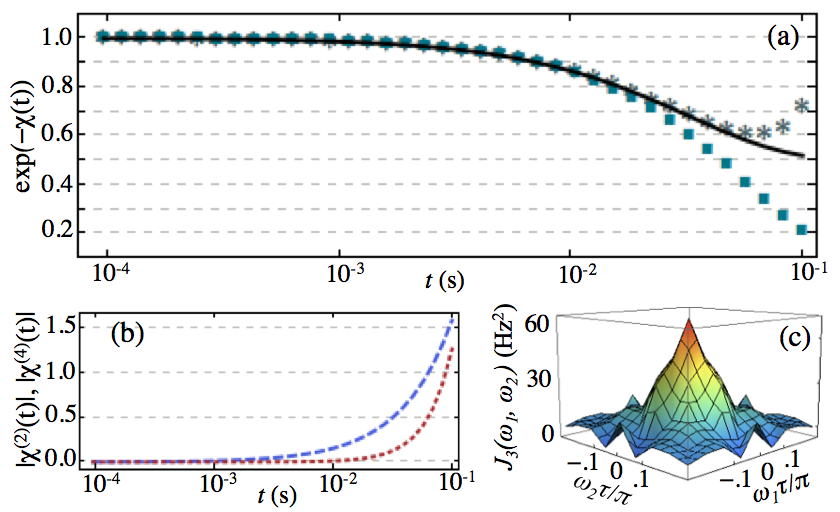}
\caption{(Color online) Qubit decay under non-Gaussian spin-boson dephasing (a), relative strengths of the first two 
terms in the cumulant expansion for $\chi(t)$ (b), reconstructed effective tri-spectrum (c). 
The non-Gaussian initial bath state is $\rho_B(0)=\rho_{T_1}/2+\rho_{T_2}/2$, where $\rho_{T_1}$, $\rho_{T_2}$ are 
thermal states at temperatures $T_1=7.64$K, $T_2=7.64\times 10^3$K. Ohmic spectral density 
$J(\omega)=w_0|\omega/\omega_c|e^{-(\omega/\omega_c)^2}$ is assumed, with 
$w_0=0.1$ nHz, $\omega_c=10$ kHz. The curves in (a) represent theoretical decay (black solid), decay predicted by 
reconstructing $S_\text{eff}(\omega)$ and $J_3(\omega_1, \omega_2)$ (grey asterisks), and decay predicted by 
approximating the noise as Gaussian and reconstructing $S_\text{eff}(\omega)$ only (teal squares). 
The non-Gaussian prediction in (b) fails when $|\chi^{(4)}(t)|$ and $|\chi^{(2)}(t)|$ become comparable. 
All reconstructions used $M=50$ repetitions of 21 
base sequences composed of CDD$_{0-5}$, with $\tau=3.44\times 10^{-5}\,$s, $T=32\tau$. 
}
\label{fig::SB}
\end{figure}

For illustration, we choose here $\rho_B(0)$ to be a non-Gaussian, non-separable state corresponding to 
far-from-equilibrium conditions, and simultaneously reconstruct $S_\text{eff}(\omega)$ and $J_3(\omega_1, \omega_2)$ 
by numerically determining $\chi_{[p]^M}$ 
and inverting the appropriate system of linear equations \cite{Supplement}.
To test the accuracy of our reconstructions, we again predict the dynamics of the qubit under free evolution.  
As shown in Fig. \ref{fig::SB}(a),  
taking into account the non-Gaussianity of the noise by reconstructing both the effective spectrum and tri-spectrum 
improves the prediction by almost an order of magnitude in time. 
Because the non-Gaussian prediction only uses spectral quantities associated with the second and fourth cumulants, 
however, it fails when the fourth cumulant becomes comparable in size to the second, indicating that the higher-order 
cumulants can no longer be neglected (see also Fig. \ref{fig::SB}(b)).

{\em Conclusion.--} We introduced control protocols for characterizing the high-order spectra associated with 
non-Gaussian dephasing on a qubit probe coupled to a classical or a quantum bosonic environment.  
Our approach overcomes limitations of existing protocols, allowing in 
particular for spectral reconstruction over an extended bandwidth, which is of independent 
interest for quantum sensing applications.  
Our work also points to the need for a deeper understanding of high-order quantum noise spectra 
-- beginning from more complex dephasing settings described by non-linear spin-boson models or spin baths.  
We expect implementation of our protocols to be within reach for various device technologies, 
in particular transmon or flux qubits \cite{Oliver}, where they may help shed light onto the microscopic 
origin of the noise itself. 

We thank M. Biercuk, W. Oliver, S. Gustavsson, and F. Yan for valuable discussions.
Work at Dartmouth was supported from the US ARO under contract No. W911NF-14-1-0682 
and the Constance and Walter Burke Special Projects Fund in Quantum Information Science.
GAPS acknowledges support from the ARC Centre of Excellence grant No. CE110001027.


%



\onecolumngrid
\appendix
\section{SUPPLEMENTARY INFORMATION}

\subsection{I. Expanded bandwidth of spectral reconstructions}
\label{sec::HighFreq}

The range $|\omega|\leq 2\pi/T_{\text{min}}= \pi/\tau$ is commonly regarded as a bound in spectral reconstructions. 
Rather than being a fundamental limit, however, this bound only holds for certain classes of control sequences.  Here, we show that the ultimate bound is set by the {\em time resolution} of the control sequence, $\delta$. Specifically,
\begin{align}
|\omega|\leq\pi/\delta.
\end{align}
This bound is saturated provided this set of sequences satisfies certain conditions, which we outline below.

Consider a set of base control sequences $P$, with sequence $p\in P$ having duration $T_p=T/n_p$, where $T=\text{max}\{T_p|\,p\in P\}$ and $n_p\in\mathbb{Z}^{+}$. For example, all $n_p=1$ and $T_p=T$ for the family we construct from different orders of CDD, whereas  the $n_p$ are different positive integers and the $T_p$ are rational multiples of $T$ for the CPMG sequences used in the protocol of Alvarez and Suter. Each sequence $p$ has $N_p$ pulses at times $\{t_1,\ldots,t_{N_p}\}$. Define $t_0=0$ and $t_{N_p+1}=T_p$. In practice, the times  
$\{ t_0,\ldots, t_{N_p+1}\}$ are constrained by the {time resolution} $\delta$: since times less than $\delta$ are not resolvable, all times must be integer multiples of $\delta$, say $t_i=z_i\delta$ for $z_i\in\mathbb{N}$. Consequently, the time separations between any two times $t_i$ and $t_j$ must also be integer multiples of $\delta$. We denote the time separations by $|t_i-t_j|=z_{i,j}^p\delta$ for $z_{i,j}^p\in\mathbb{Z}^+$, $i,j\in\{ 0,\ldots, N_p+1\}, \, i\neq j$.

For simplicity, consider the task of reconstructing the spectrum of a stationary Gaussian noise process. This requires measuring $\chi_{[p]^M}$, the decay parameter after $M$ repetitions of the control sequence $p$. \erf{eq::chiM} in the main text enables us to relate $\chi_{[p]^M}$ to the PSD evaluated at a set of harmonic frequencies $\{0,2\pi/T_p...,2\pi h_p/T_p\}$ for $h_p\in\mathbb{Z}^+$. In the Gaussian case,  \erf{eq::chiM} reduces to
\begin{align}
\chi_{[p]^M}\approx &
-\frac{ M}{2T_p}\sum_{h=0}^{h_p} m_{1}\!\!\left(\frac{2\pi h}{T_p}\right)
\Big|F_p\!\left(\frac{2\pi h}{T_p}\right)\Big|^2S\!\left(\frac{2\pi h}{T_p}\right)\notag \\
=
&
-\frac{M}{2T_p}\sum_{h=0}^{h_p} m_{1}\!\!\left(\frac{2\pi n_ph}{T}\right)
\Big|F_{p}\!\left(\frac{2\pi n_ph}{T}\right)\Big|^2S\!\left(\frac{2\pi n_ph}{T}\right)
\label{eq::chiM2}.
\end{align}
Inverting the system of linear equations formed by \erf{eq::chiM2} for each $p$ in $P$ enables one to reconstruct the PSD at the ``base harmonics", integer multiples of $2\pi/T$. 

Non-degeneracy of this linear system depends on the periodicity of the FFs. The periodicity of sequence $p$ is determined by the oscillatory component of $|F_p(\omega)|^2$, which is given by
 \begin{align}
O_p(\omega)\equiv &|\omega F_p(\omega)|^2
\\=&2+4N_p+4\!\!\!\!\sum_{j, k=1,\, j\neq k}^{N_p}\!\!\!\!(-1)^{j+k}\text{cos}[(t_k-t_j)\omega]-2(-1)^{N_p}\text{cos}[(t_{N_p+1}-t_0)\omega]\label{eq::FFnumerator}\\
 &-4\sum_{j=1}^{N_p}(-1)^{j+N_p}\text{cos}[(t_j-t_{N_p+1})\omega]+4\sum_{j=1}^{N_p}(-1)^j\text{cos}[(t_j-t_0)\omega]\notag.
 \end{align}
Note that the last two terms in this expression cancel when $p$ is a time-symmetric sequence. In an abuse of notation, define $\text{gcf}\{a_1,\ldots,a_n\} \equiv \text{max}\{a\in\mathbb{R}|\,a_i/a\in\mathbb{Z}^+\,\forall i\}$, and $\text{lcm}\{a_1,\ldots ,a_n\}=\text{min}\{a\in\mathbb{R}|\,a/a_i\in\mathbb{Z}^+\,\forall i\}$. For integer $a_1,\ldots ,a_n$, these definitions reduce to the ordinary gcf and lcm, respectively. 
In the frequency domain, the periodicities of the cosine terms in \erf{eq::FFnumerator} are determined by the  time separations $t_i-t_j$. 
From these cosine terms, the periodicity of  $O_p(\omega)$ is  $\Omega^p=2\pi/\Delta t_p$, where
\begin{align}
\Delta t_p\equiv&\text{gcf}\big(\big\{|t_i-t_j|\big|\,i,j\!\in\!\{n_{-}^p,...,n_{+}^p\} \!\big\}\!\cup\!\{|t_0-t_{N_p+1}|\}\big)
\\=&\delta\,\text{gcf}\big(\big\{z_{i,j}^p|\,i,j\!\in\!\{n_{-}^p,...,n_{+}^p\} \!\big\}\!\cup\!\{z_{0,N_p+1}^p\}\big).
\end{align}
Here, $\{n_{-}^p,n_{+}^p\}=\{1,N_p\}$ when $p$ is time-symmetric and $\{n_{-}^p,n_{+}^p\}=\{0,N_p+1\}$ otherwise. Note that the arguments of the cosine terms in \erf{eq::FFnumerator} are even multiples of $\pi$ when $\omega=\Omega^p$. Consider now the ``half periodicity", 
$\omega_{\text{max}}^p=\pi/\Delta t_p$. As expected, the oscillatory component is symmetric about $\omega_{\text{max}}^p$, i.e. $O_p(\omega_{\text{max}}^p-\omega)=O_p(\omega_{\text{max}}^p+\omega)$ for all $\omega$. 

We now extend this notion of periodicity to a set of control sequences $P$. Define the half periodicity of $P$ as 
\begin{align}
\omega_{\text{max}}^P\equiv\text{lcm}\{\omega_{\text{max}}^p|\,p\in P\}=\pi/\Delta t_P, 
\end{align}
where
\begin{align}
\Delta t_P\equiv&\text{gcf}\big\{\Delta t_p\,|\, p\in P  \big \}\label{eq::maxgcfdelt}
\\=&\delta\,\text{gcf}\Big(\big\{z_{i,j}^p|\,i,j\!\in\!\{n_{-}^p,...,n_{+}^p\},p\in P \big\}\!\cup\!\{z_{0,N_p+1}^p|\,p\in P\}\Big)\label{eq::maxgcf}.
\end{align}
Note that   $O_p(\omega_{\text{max}}^P-\omega)=O_p(\omega_{\text{max}}^P+\omega)$ for all $\omega$ and $p\in P$. Because $\Delta t_P$ divides all time separations by an integer, $T=q_P\Delta t_P$ for some $q_P\in\mathbb{Z}^+$, implying $\omega_{\text{max}}^P= q_P\pi/T$. In the case that $q_P$ is even, $\omega_{\text{max}}^P\pm 2\pi/T=(q_P\pm 2)\pi/T$ are base harmonics. This implies $O_p((q_P+2)\pi/T)=O_p((q_P-2)\pi/T)$ and 
\begin{align}
|F_{p}((q_P+2)\pi/T)|^2=\left(\frac{q_P-2}{q_P+2}\right)^2|F_{p}((q_P-2)\pi/T)|^2,\; \forall\,p.
\end{align}
Consequently, the set of linear equations formed by \erf{eq::chiM2} for each $p\in P$ is degenerate and non-invertible when $\omega>\omega_{\text{max}}^P$. Similarly, if $q_P$ is odd, then $\omega_{\text{max}}^P\pm \pi/T=(q_P\pm 1)\pi/T$ are base harmonics and 
\begin{align}
|F_{p}((q_P+1)\pi/T)|^2=\left(\frac{q_P-1}{q_P+1}\right)^2|F_{p}((q_P-1)\pi/T)|^2,\; \forall\,p.
\end{align}
for all $p$, also making the set of linear equations formed by \erf{eq::chiM2} for each $p\in P$ degenerate and non-invertible. In both cases, it is impossible to reconstruct the spectrum beyond $\omega_{\text{max}}^P$.

Degeneracy of the linear inversion prohibits reconstruction beyond $|\omega|\leq\omega_{\text{max}}^P$. More generally, for \erf{eq::chiM2} to be invertible in the range $|\omega|\leq\omega_{\text{max}}^P$, $P$ must contain a {\em sufficiently large} number of control sequences. If $2\pi h_{\text{max}}/T$ is the greatest base harmonic less than $\omega_{\text{max}}^P$, reconstructing the harmonics $\{2\pi/T,\dots,2\pi h_{\text{max}}/T\}$ requires {\em at least} $h_{\text{max}}$ control sequences. (Alternatively, when using sequences with zero-filtering order, reconstructing the harmonics $\{0,\ldots,2\pi h_{\text{max}}/T\}$ requires at least $h_{\text{max}}+1$ control sequences.) Provided that there are a sufficient number of control sequences, the bound on spectral reconstruction is set by the maximum possible $\omega_{\text{max}}^P$. When the gcf in  \erf{eq::maxgcf} is equal to one,  $\Delta t_P$ is minimal and $\omega_{\text{max}}^P=\pi/\Delta t_P$ is maximal. The half periodicity is, thus, bound by
\begin{align}\label{eq::resbound}
\omega_{\text{max}}^P\leq\pi/\delta.
\end{align}
This bound is saturated when there exist time separations in $P$ that are mutually prime integer multiples of $\delta$. Alternatively, from $\omega_{\text{max}}^P=\text{lcm}\{\omega_{\text{max}}^p|\,p\in P\}$ we see that $\omega_{\text{max}}^P$ is maximized when $P$ contains sequences with {\em incommensurate periodicities}.

{\bf Examples.--} (1) Consider first a set of {\em digital} sequences, where the minimal time between adjacent pulses is $\tau$ and  the pulses in all sequences are spaced by integer multiples of $\tau$. From \erf{eq::maxgcfdelt}, $\Delta t_P=\tau$, prohibiting spectral reconstruction 
beyond $|\omega|\leq\pi/\tau$. 

(2) As mentioned in the text, the protocol of Alvarez and Suter uses CPMG sequences with varying cycle times $T,\,T/2,...,T/n=2\tau$ 
for some $n\in\mathbb{Z}^+$. 
A CPMG sequence $p$ with cycle time $T/n_p=2n\tau/n_p$ has two pulses at times $t_1=n\tau/(2n_p)$ and $t_2=3n\tau/(2n_p)$. 
Because CPMG is time-symmetric, 
\begin{align}
\Delta t_P=\text{gcf}\{n\tau/n_p|\,p\in P\}
=n\tau\,\text{gcf}\{1/n_p|\,p\in P\}
=\frac{n\tau}{\text{lcm}\{n_p|\,p\in P\}}.
\end{align}
Provided that $n<\text{lcm}\{n_p|\,p\in P\}$, $\Delta t_P<\tau$ and $\omega_{\text{max}}^P>\pi/\tau$. In practice, however, the range of spectral reconstruction can never go beyond $|\omega|\leq\pi/\tau$ even when $\omega_{\text{max}}^P>\pi/\tau$. This occurs because using $n$ CPMG sequences with cycle times $T,\,T/2,...,T/n=2\tau$ allows one to reconstruct at most $n$ base harmonics. The maximal  harmonic that can be reconstructed is, therefore, $\omega=2\pi n/T=2\pi n/(n\tau)=\pi/\tau$. The reconstruction is, thus, bound by $|\omega|\leq\pi/\tau$.

(3) The sequences we construct by combining different orders of CDD permit spectral reconstructions beyond $|\omega|\leq\pi/\tau$ and {\em can saturate the bound} in \erf{eq::resbound}. All sequences have a fixed cycle time $T=q\delta$ for some $q\in\mathbb{Z}^+$. We create our sequences by partitioning $q$ into integers $q_i$ such that $q=\sum_iq_i$. Subject to the constraint that the time between two adjacent pulses is no smaller than $\tau$,
we place a CDD$_d$ sequence into the subinterval of total time duration $q_i\delta$. Whereas the protocol of Alvarez and Suter allows for only $n$ sequences for a fixed cycle time $T=2n\tau$, our procedure allows for more sequences. This enables us to reconstruct the spectrum beyond $|\omega|\leq\pi/\tau$ for a sufficiently large $\omega_{\text{max}}^P$. Our construction procedure also permits us to reach the ultimate bound set by the pulse timing resolution, $\omega_{\text{max}}^P=\pi/\delta$. Consider a subset of sequences $P$ drawn from our family of sequences. Suppose that the sequence $p\in P$ has a subinterval of total time $q_1\delta=2m\delta$ for some $m\in\mathbb{Z}^+$. Into this subinterval, we can place a CDD$_1$ sequence with a time $m\delta$ between the pulses. (CDD$_1$ over a time interval $t$ consists of two pulses  at  $t_1=t/2$ and $t_2=t$). Suppose that either $p$ or another sequence in $P$ contains a subinterval of total time $q_2\delta=2(m+1)\delta$. Into this subinterval, we can place a CDD$_1$ sequence with a time $(m+1)\delta$ between the pulses. Because $\text{gcf}\{(m+1)\delta,m\delta\}=\delta$, $\Delta t_P=\delta$ regardless of the other sequences in $P$. Thus, $\omega_{\text{max}}^P=\pi/\delta$ and the bound is saturated. 

In general, the expanded spectral range may come at a cost of decreased overall numerical stability, as the conditioning 
of the linear inversion at fixed $\delta$, $\tau$ and $T$ worsens with an increasing number of harmonics -- which may be 
a practical limiting factor in pushing to very high frequencies.
In Fig. \ref{fig::Lorentzian3PiTau}, the bound of spectral reconstruction is extended to $|\omega|\leq 3\pi/\tau$ by using $\delta<\tau/3$. 
Because of the large number of sequences that can be generated for a fixed $\delta$, we used a random search to select 25 control 
sequences that produced a well conditioned linear inversion.

\subsection{II. Leading order cumulants and spectra of the linear spin-boson model}
\label{sec::SB}

Here, we derive the spectrum and tri-spectrum of the stationary linear spin boson model from the second and fourth cumulants, respectively. 

{\bf Leading order cumulants.--}
In the interaction picture with respect to the bath Hamiltonian, $H_B=\hbar\sum_k\Omega_ka_k^\dag a_k$, the qubit couples to the bath 
operator $B(t)=\sum_k(g_ke^{i\Omega_kt}a_k^\dag+g_k^{*}e^{-i\Omega_kt}a_k)$. The quantum noise cumulants depend on moments of the bath operator $B(t)$ taken with respect to the initial bath state, i.e., $\expect{B(t_1)...B(t_n)}=\text{Tr}[\rho_B(0)B(t_1)...B(t_n)]$. The noise is stationary if and only if $[\rho_B(0), H_B]=0$. Because initial bath states satisfying this condition are diagonal in the multimode Fock basis, all odd moments (and consequently odd cumulants) are zero. The first two leading order cumulants are, thus, the second and the fourth, where a nonzero fourth cumulant is the leading order signature of non-Gaussianity. For a stationary initial bath state, they are given by
\begin{align}
C^{(2)}(t_1,t_2)=\expect{B(t_1)B(t_2)}=\sum_k|g_k|^2\Big(e^{i\Omega_k(t_2-t_1)}\expect{n_k+1}+e^{-i\Omega_k(t_2-t_1)}\expect{n_k}\Big)\label{eq::SBC2}.
\end{align}
and
\begin{align}
C^{(4)}(t_1,t_2,t_3,t_4)&=\expect{B(t_1)B(t_2)B(t_3)B(t_4)}-\expect{B(t_1)B(t_2)}\expect{B(t_3)B(t_4)}
-\expect{B(t_1)B(t_3)}\expect{B(t_2)B(t_4)}\notag\\&\,\,\,\,\,\,\,-\expect{B(t_1)B(t_4)}\expect{B(t_2)B(t_3)}\notag\\
&=\sum_{k,l}\sum_{a,b=\pm 1}|g_k|^2|g_l|^2
\Big[\Big(\expect{n_kn_l}-\expect{n_k}\expect{n_l}\Big)_{k\neq l}+\delta_{kl}\frac{1}{2}\Big(\expect{n_k^2}-2\expect{n_k}^2-\expect{n_k}\Big)\Big]\label{eq::SBC4}\\\notag
&\,\,\,\,\,\,\,\times\Big(e^{i\Omega_ka(t_1-t_2)}e^{i\Omega_lb(t_3-t_4)}+
e^{i\Omega_ka(t_1-t_3)}e^{i\Omega_lb(t_4-t_2)}+e^{i\Omega_ka(t_1-t_4)}e^{i\Omega_lb(t_2-t_3)}\Big).
\end{align}
Note that when the initial bath state is separable, $\big(\expect{n_kn_l}-\expect{n_k}\expect{n_l}\big)_{k\neq l}=0$ for all $k\neq l$. This implies that $C^{(4)}(t_1,t_2,t_3,t_4)$ depends on a single mode rather than two.

{\bf Spectrum.--}
For a stationary initial bath state, the spectrum is the Fourier transform of $C^{(2)}(t_1,t_2)$ in \erf{eq::SBC2} with 
respect to the time separation $\tau=t_2-t_1$, 
\begin{align}
S(\omega)=2\pi\sum_k|g_k|^2\Big(\delta(\omega-\Omega_k)\expect{n_k+1}+\delta(\omega+\Omega_k)\expect{n_k}\Big).\label{eq::Sdiscrete}
\end{align}
Note that $S(\omega)$ is a discontinuous function of $\omega$ as long as the energy levels of the bath are discrete. In the (standard) 
limit where the spacing between the bath energy levels is small, however, we can treat the bath as a continuum. In this regime, 
the spectrum becomes
\begin{align}
S(\omega)=\bigg\{\begin{array}{ccc}2\pi J(|\omega|)(\mathcal{N}(|\omega|)+1),&&\omega>0,
\\2\pi J(|\omega|)\mathcal{N}(|\omega|),&&\omega<0,\end{array}
\end{align}
where $\mathcal{N}(\omega)=\expect{n_k}|_{\Omega_k=\omega}$ and $J(\omega)=\sum_k|g_k|^2[\delta(\omega-\Omega_k)+\delta(\omega+\Omega_k)]$ is the spectral density function of the bath. On account of its asymmetry about $\omega=0$, $S(\omega)$ cannot be reconstructed by our protocol. Because the filter functions of the control sequences are symmetric about $\omega=0$, reconstructions are limited to the symmetric component of $S(\omega)$. This symmetric component, which we term the effective spectrum, is given by
\begin{align}
S_{\text{eff}}(\omega)=\frac{1}{2}\big[S(\omega)+S(-\omega)\big]=\pi J(|\omega|)(2\mathcal{N}(|\omega|)+1).
\end{align}
Reconstruction of $S_{\text{eff}}(\omega)$ will be treated in detail in the following section.

{\bf Trispectrum.--} The trispectrum of the stationary linear spin boson model takes different forms depending on whether the initial bath state is separable. We first consider the case where there is entanglement between different modes of the initial bath state, i.e. there exists different modes $k$ and $l$ such that $\big(\expect{n_kn_l}-\expect{n_k}\expect{n_l}\big)_{k\neq l}$ in \erf{eq::SBC4} is nonzero. By taking the Fourier transform of \erf{eq::SBC4} with respect to the time separations $\tau_i=t_{i+1}-t_1$ for $i\in\{1,2,3\}$, we obtain the trispectrum,
\begin{align}
S_3(\vec{\omega}_3)=&(2\pi)^3\sum_{k,l}\sum_{a,b=\pm 1}|g_k|^2|g_l|^2\Big[\Big(\expect{n_kn_l}-\expect{n_k}\expect{n_l}\Big)_{k\neq l}+\delta_{kl}\frac{1}{2}\Big(\expect{n_k^2}-2\expect{n_k}^2-\expect{n_k}\Big)\Big]\label{eq::Tdiscrete}\\
&\,\,\,\,\,\,\,\times\Big(\delta(\omega_1-a\Omega_k)\delta(\omega_2+a\Omega_k)\delta(\omega_3-b\Omega_l)
+\delta(\omega_1-a\Omega_k)\delta(\omega_3+a\Omega_k)\delta(\omega_2+b\Omega_l)\notag
\\&\,\,\,\,\,\,\,\,\,\,\,\,\,+\delta(\omega_1-a\Omega_k)\delta(\omega_2-b\Omega_l)\delta(\omega_3+b\Omega_l)\Big)\notag
\end{align}
Once again, we make the continuum approximation on the bath and the trispectrum becomes
\begin{align}
\label{eq::sbTri}
S_3(\vec{\omega}_3)=&(2\pi)^3\Big[\delta(\omega_1+\omega_2)J_3(\omega_1,\omega_3)+
\delta(\omega_1+\omega_3)J_3(\omega_2,\omega_3)+
\delta(\omega_2+\omega_3)J_3(\omega_1,\omega_2)\Big],
\end{align}
where 
\begin{align}
J_3(\vec{\nu}_2)=J(|\nu_1|)J(|\nu_2|)N(|\nu_1|,|\nu_2|),
\end{align}
is the effective trispectrum,
\begin{align}
N(\vec{\nu}_2)=\Big[\Big(\expect{n_kn_l}-\expect{n_k}\expect{n_l}\Big)_{k\neq l}+\delta_{kl}\frac{1}{2}\Big(\expect{n_k^2}-2\expect{n_k}^2-\expect{n_k}\Big)\Big]_{(\Omega_k=\nu_1,\,\Omega_l=\nu_2)}. 
\end{align}
and $J(\nu)$ is the spectral density. Unlike the spectrum, delta functions remain in the trispectrum after making the continuum approximation. As detailed in the following section, our protocol cannot reconstruct the trispectrum because of these delta function discontinuities. 
Provided it is smooth, we can reconstruct the effective trispectrum.  Precisely characterizing what class of states $\rho$ results in smooth 
noise polyspectra is an interesting question, which has not been addressed to the best of our knowledge and we leave to future investigation.

We can similarly examine the trispectrum of an initially separable bath state. Separability implies $\big(\expect{n_kn_l}-\expect{n_k}\expect{n_l}\big)_{k\neq l}= 0$ for all $k\neq l$ in \erf{eq::SBC4}. The fourth cumulant, which depends on a single bath mode, reduces to
\begin{align}
C^{(4)}(t_1,t_2,t_3,t_4)&=\sum_{k}\sum_{a=\pm 1}|g_k|^4
\Big(\expect{n_k^2}-2\expect{n_k}^2-\expect{n_k}\Big)\label{eq::SBC4sep}\\\notag
&\,\,\,\,\,\,\,\times\Big(e^{i\Omega_ka(t_1-t_2+t_3-t_4)}+
e^{i\Omega_ka(t_1-t_3+t_4-t_2)}+e^{i\Omega_ka(t_1-t_4+t_2-t_3)}\Big).
\end{align}
To determine the trispectrum, we  take the Fourier transform of \erf{eq::SBC4sep} with respect to the time separations $\tau_i=t_{i+1}-t_1$ for $i\in\{1,2,3\}$,
\begin{align}
S_3(\vec{\omega}_3)=&\sum_k\sum_{a=\pm1}|g_k|^4\Big(\expect{n_k^2}-2\expect{n_k}^2-\expect{n_k}\Big)
\Big(\delta(\omega_1-a\Omega_k)\delta(\omega_2+a\Omega_k)\delta(\omega_3-a\Omega_k)
\\&\,\,\,\,\,\,\,+\delta(\omega_1-a\Omega_k)\delta(\omega_2+a\Omega_k)\delta(\omega_3+a\Omega_k)
+\delta(\omega_1-a\Omega_k)\delta(\omega_2-a\Omega_k)\delta(\omega_3+a\Omega_k)\Big)\notag.
\end{align}
After making the continuum approximation on the bath, this expression becomes
\begin{align}\label{eq::sbTriSep}
S_3(\vec{\omega}_3)=&(2\pi)^3\big[\delta(\omega_1+\omega_2)\delta(\omega_1+\omega_3)j_3(\omega_1)+
\delta(\omega_1+\omega_2)\delta(\omega_2+\omega_3)j_3(\omega_2)\\&\,\,\,\,\,
+\delta(\omega_1+\omega_3)\delta(\omega_2+\omega_3)j_3(\omega_3)\big],\notag
\end{align}
where 
\begin{align}
j_3(\omega)=J(|\omega|)^2n(|\omega|)
\end{align}
is the effective tri-spectrum for the separable case, $n(\omega)=\big(\expect{n_k^2}-2\expect{n_k}^2-\expect{n_k}\big)|_{\Omega_k=\omega}$ and $J(\omega)$ is the spectral density. Like the trispectrum of the entangled case, $S_3(\vec{\omega}_3)$ exhibits delta function discontinuities in the separable case that persist in the continuum limit. As in the entangled case, $j_3(\omega)$ can be reconstructed provided it is smooth.

\subsection{III. Spectroscopy with the linear spin-boson model}
\label{sec::SBspec}

Here, we show how our protocol can reconstruct the effective spectrum and tri-spectrum from the leading order cumulants of the linear spin boson model.  Due to the asymmetry of the spectrum and the discontinuities present in the tri-spectrum, reconstruction of the leading order polyspectra for the linear spin boson model differs from the classical case in a few respects. As in the classical case, we use repetition of control sequences to generate a frequency comb that allows us to relate the decay parameter to values of the effective spectrum and tri-spectrum at the harmonic frequencies. While the effective spectrum is not equivalent to the actual spectrum, the system of linear equations relating $S_{\text{eff}}(\omega)$ to $\chi_{[p]^M}$ is identical to the classical case. Due to the delta functions present in the tri-spectrum, however, the system of linear equations relating $J_{\text{eff}}(\omega)$ to $\chi_{[p]^M}$ has a different dependence on $M$ than in the classical case. Consequently, the form of the inversion we use to reconstruct the effective tri-spectrum is specific to the linear spin boson model. It should be noted that this inversion is still widely applicable, however, as noise in a variety of physical and quantum information settings is modeled using a linearly coupled bosonic bath. 

Deriving the system of linear equations needed to reconstruct the effective spectrum and tri-spectrum begins with the cumulant expansion 
for the decay parameter $\chi(t)$, given in the main text. At small times, such expansion can be truncated at $\ell=2$, leaving two terms $\chi^{(2)}(t)$ and $\chi^{(4)}(t)$ corresponding to $\ell=1$ and $\ell=2$, respectively.  Using Eq. (\ref{eq::FTcumulantM}), 
$\chi^{(2)}$ and $\chi^{(4)}$ after $M$ repetitions of a control sequence $p$ are given by
\begin{align}\label{eq::chi2}
\chi_{[p]^M}^{(2)}=&\frac{1}{4\pi}\int_{-\infty}^{\infty}\!\!\!\!\!\!d\omega|F_p(\omega)|^2\frac{\text{sin}^2(M\omega T/2)}{\text{sin}^2(\omega T/2)}S(\omega),
\end{align}
and
\begin{align}
\label{eq::chi4}
\chi_{[p]^M}^{(4)}=
\frac{1}{4!(2\pi)^3}\!\!\!
\int_{{\mathbb R}^3} d\vec{\omega}_3
\prod_{j=1}^3\!\!\Big[F_p(\omega_j)\frac{\text{sin}(M\omega_j T/2)}{\text{sin}(\omega_j T/2)}\Big]F_p[-(\omega_1+\omega_2+\omega_3)]
\frac{\text{sin}[M(\omega_1+\omega_2+\omega_3) T/2]}{\text{sin}[(\omega_1+\omega_2+\omega_3)T/2]}S_3(\vec{\omega}_3).
\end{align}

{\bf Effective spectrum equations.--} From $\chi_{[p]^M}^{(2)}$, we derive the system of linear equations for the effective spectrum. Provided that $S(\omega)$ is smooth, the comb approximation in \erf{eq::hypercomb} holds. Substituting into \erf{eq::chi2} produces
\begin{align}
\chi_{[p]^M}^{(2)}&=\frac{M}{2T}\sum_{k=-\infty}^{\infty}\Big|F_p\Big(\frac{2\pi k}{T}\Big)\Big|^2S\Big(\frac{2\pi k}{T}\Big)\\\label{eq::chiSeff}
&=\frac{M}{2T}|F_p(0)|^2S_{\text{eff}}(0)+\frac{M}{T}\sum_{k=1}^{K}\Big|F_p\Big(\frac{2\pi k}{T}\Big)\Big|^2S_{\text{eff}}\Big(\frac{2\pi k}{T}\Big),
\end{align}
where in the second equality we use $|F_p(\omega)|^2=|F_p(-\omega)|^2$ and truncate the sum at a finite $k=K$. Note the form of this linear equation is identical to the $\ell=1$ term of \erf{eq::chiM}, restricted to the set of nonnegative harmonics where $m_1(0)=1$ and $m_1(\omega>0)=2$. 

{\bf Effective tri-spectrum equations.--} Reconstructing the effective trispectrum requires $\chi_{[p]^M}^{(4)}$. This reconstruction is specialized to the spin-boson case, where $S_3(\vec{\omega}_3)$ takes the form of \erf{eq::sbTri} when the initial bath state is entangled and \erf{eq::sbTriSep} when the initial bath state is separable. 

We treat  the entangled case by subsituting \erf{eq::sbTri} into \erf{eq::chi4} and using permutation symmetry of the frequency coordinates to obtain
\begin{align}\label{eq::chi4Sep}
\chi_{[p]^M}^{(4)}=&\frac{1}{8}\int_{-\infty}^{\infty}\!\!\!\!\!\!\!\!d\omega_1\int_{-\infty}^{\infty}\!\!\!\!\!\!\!\!d\omega_2|F_p(\omega_1)|^2|F_p(\omega_2)|^2
\frac{\text{sin}^2(M\omega_1 T/2)}{\text{sin}^2(\omega_1 T/2)}\frac{\text{sin}^2(M\omega_2 T/2)}{\text{sin}^2(\omega_2 T/2)}
J_3(\omega_1,\omega_2).
\end{align}
If $J_3(\omega_1,\omega_2)$ is smooth and $M\gg 1$, repetition produces a hyper-comb
\begin{align}\label{eq::combSB}
\frac{\text{sin}^2(M\omega_1 T/2)}{\text{sin}^2(\omega_1 T/2)}\frac{\text{sin}^2(M\omega_2 T/2)}{\text{sin}^2(\omega_2 T/2)}\approx \Big(\frac{2\pi M}{T}\Big)^2
\sum_{n_1=-\infty}^{\infty}\sum_{n_2=-\infty}^{\infty}\delta \Big(\omega_1-\frac{2\pi n_1}{T}\Big)\delta \Big(\omega_2-\frac{2\pi n_2}{T}\Big).
\end{align}
By substituting the hyper-comb into \erf{eq::chi4Sep}, we obtain a linear equation relating $\chi_{[p]^M}^{(4)}$ to the effective trispectrum,
\begin{align}\label{eq::chi4SepLin}
\chi_{[p]^M}^{(4)}=&\frac{1}{2}\Big(\frac{\pi M}{T}\Big)^2 \sum_{(h_1,h_2)\in O_2}m_2(h_1,h_2)|F_p(h_1)|^2|F_p(h_2)|^2
J_3(h_1,h_2).
\end{align}
Here, $O_2$ is a subset of harmonics contained in what is effectively the principle domain of $J_3(\omega_1,\omega_2)$, an octant in $\mathbb{R}^2$. That the effective trispectrum can be fully specified by knowing its value on an octant is a consequence of $J_3(\omega_1,\omega_2)$ being invariant under sign flips and permutations of $\omega_1$ and $\omega_2$. In \erf{eq::chi4SepLin}, $m_2(\omega_1,\omega_2)$ is a multiplicity equal to the number of points where the effective trispectrum has the value $J_3(\omega_1,\omega_2)$ for $(\omega_1,\omega_2)\in O_2$.

For the separable case, we substitute  \erf{eq::sbTriSep} into \erf{eq::chi4} and again use permutation symmetry of the frequency coordinates to obtain
\begin{align}\label{eq::chi4Sep2}
\chi_{[p]^M}^{(4)}=&\frac{1}{8}\int_{-\infty}^{\infty}\!\!\!\!\!\!\!\!d\omega_1|F_p(\omega_1)|^4
\frac{\text{sin}^4(M\omega_1 T/2)}{\text{sin}^4(\omega_1 T/2)}j_3(\omega_1).
\end{align}
If $j_3(\omega_1)$ is smooth and $M\gg 1$, we can make the frequency comb approximation 
\begin{align}
\frac{\text{sin}^4(M\omega_1 T/2)}{\text{sin}^4(\omega_1 T/2)}\approx\frac{4\pi M^3}{3T}\sum_{k=-\infty}^\infty\delta\Big(\omega_1-\frac{2\pi k}{T}\Big).
\end{align}
From the frequency comb, we again obtain a linear equation relating $\chi_{[p]^M}^{(4)}$ to the effective trispectrum
\begin{align}\label{eq::chi4Sep2Lin}
\chi_{[p]^M}^{(4)}=&\frac{\pi M^3}{6T}|F_p(0)|^4j_3(0)+\frac{\pi M^3}{3T}\sum_{k=1}^K\Big|F_p\Big(\frac{2\pi k}{T}\Big)\Big|^4j_3\Big(\frac{2\pi k}{T}\Big).
\end{align}
Here, we have truncated the sum to a finite $k=K$.

{\bf Reconstructing the effective spectra.--} Because the effective spectrum and tri-spectrum are both associated with even cumulants, we reconstruct these quantities simultaneously by measuring the decay parameters $\chi_{[p]^M}$ associated with a set of control sequences.  From the expression for $\chi_{[p]^M}^{(2)}$ in \erf{eq::chiSeff} and the expressions for $\chi_{[p]^M}^{(4)}$ in \erf{eq::chi4SepLin} or \erf{eq::chi4Sep2Lin}, we can relate $\chi_{[p]^M}$ to the effective spectrum and trispectrum evaluated at the harmonic frequencies.  For simplicity, we assume prior knowledge about the bath and use the form of $\chi_{[p]^M}^{(4)}$ in \erf{eq::chi4SepLin}, which we also used in the reconstructions presented in the text.   From $\chi_{[p]^M}=\chi_{[p]^M}^{(2)}+\chi_{[p]^M}^{(4)}$,  we obtain
\begin{align}\label{eq::SBChi}
\chi_{[p]^M}=&\frac{M}{2T}|F_p(0)|^2S_{\text{eff}}(0)+\frac{M}{T}\sum_{k=1}^{K}\Big|F_p\Big(\frac{2\pi k}{T}\Big)\Big|^2S_{\text{eff}}\Big(\frac{2\pi k}{T}\Big)\\\notag&+\frac{1}{2}\Big(\frac{\pi M}{T}\Big)^2 \sum_{(h_1,h_2)\in O_2}m_2(h_1,h_2)|F_p(h_1)|^2|F_p(h_2)|^2J_3(h_1,h_2).
\end{align}
By measuring $\chi_{[p]^M}$ for a set of control sequences and inverting the system of linear equations formed by \erf{eq::SBChi}, we can obtain the effective spectrum and trispectrum.

Note that with no prior information about the noise, deciding which form of  $\chi_{[p]^M}^{(4)}$ to use will be an iterative process. For example, from a single set of measurements one can reconstruct both $J_3(\omega_1,\omega_2)$ using the entangled form and $j_3(\omega_1)$ using the separable form. These reconstructions can then be used to predict the dynamics of the qubit under free decay. By comparing the predictions to actual measurements of the qubit under free decay, one can decide which model best describes the noise, as mentioned in the text. 


\begin{thebibliography}{0}%
\makeatletter
\providecommand \@ifxundefined [1]{%
 \@ifx{#1\undefined}
}%
\providecommand \@ifnum [1]{%
 \ifnum #1\expandafter \@firstoftwo
 \else \expandafter \@secondoftwo
 \fi
}%
\providecommand \@ifx [1]{%
 \ifx #1\expandafter \@firstoftwo
 \else \expandafter \@secondoftwo
 \fi
}%
\providecommand \natexlab [1]{#1}%
\providecommand \enquote  [1]{``#1''}%
\providecommand \bibnamefont  [1]{#1}%
\providecommand \bibfnamefont [1]{#1}%
\providecommand \citenamefont [1]{#1}%
\providecommand \href@noop [0]{\@secondoftwo}%
\providecommand \href [0]{\begingroup \@sanitize@url \@href}%
\providecommand \@href[1]{\@@startlink{#1}\@@href}%
\providecommand \@@href[1]{\endgroup#1\@@endlink}%
\providecommand \@sanitize@url [0]{\catcode `\\12\catcode `\$12\catcode
  `\&12\catcode `\#12\catcode `\^12\catcode `\_12\catcode `\%12\relax}%
\providecommand \@@startlink[1]{}%
\providecommand \@@endlink[0]{}%
\providecommand \url  [0]{\begingroup\@sanitize@url \@url }%
\providecommand \@url [1]{\endgroup\@href {#1}{\urlprefix }}%
\providecommand \urlprefix  [0]{URL }%
\providecommand \Eprint [0]{\href }%
\providecommand \doibase [0]{http://dx.doi.org/}%
\providecommand \selectlanguage [0]{\@gobble}%
\providecommand \bibinfo  [0]{\@secondoftwo}%
\providecommand \bibfield  [0]{\@secondoftwo}%
\providecommand \translation [1]{[#1]}%
\providecommand \BibitemOpen [0]{}%
\providecommand \bibitemStop [0]{}%
\providecommand \bibitemNoStop [0]{.\EOS\space}%
\providecommand \EOS [0]{\spacefactor3000\relax}%
\providecommand \BibitemShut  [1]{\csname bibitem#1\endcsname}%
\let\auto@bib@innerbib\@empty
\end{thebibliography}%


\begin{thebibliography}{16}%
\makeatletter
\providecommand \@ifxundefined [1]{%
 \@ifx{#1\undefined}
}%
\providecommand \@ifnum [1]{%
 \ifnum #1\expandafter \@firstoftwo
 \else \expandafter \@secondoftwo
 \fi
}%
\providecommand \@ifx [1]{%
 \ifx #1\expandafter \@firstoftwo
 \else \expandafter \@secondoftwo
 \fi
}%
\providecommand \natexlab [1]{#1}%
\providecommand \enquote  [1]{``#1''}%
\providecommand \bibnamefont  [1]{#1}%
\providecommand \bibfnamefont [1]{#1}%
\providecommand \citenamefont [1]{#1}%
\providecommand \href@noop [0]{\@secondoftwo}%
\providecommand \href [0]{\begingroup \@sanitize@url \@href}%
\providecommand \@href[1]{\@@startlink{#1}\@@href}%
\providecommand \@@href[1]{\endgroup#1\@@endlink}%
\providecommand \@sanitize@url [0]{\catcode `\\12\catcode `\$12\catcode
  `\&12\catcode `\#12\catcode `\^12\catcode `\_12\catcode `\%12\relax}%
\providecommand \@@startlink[1]{}%
\providecommand \@@endlink[0]{}%
\providecommand \url  [0]{\begingroup\@sanitize@url \@url }%
\providecommand \@url [1]{\endgroup\@href {#1}{\urlprefix }}%
\providecommand \urlprefix  [0]{URL }%
\providecommand \Eprint [0]{\href }%
\providecommand \doibase [0]{http://dx.doi.org/}%
\providecommand \selectlanguage [0]{\@gobble}%
\providecommand \bibinfo  [0]{\@secondoftwo}%
\providecommand \bibfield  [0]{\@secondoftwo}%
\providecommand \translation [1]{[#1]}%
\providecommand \BibitemOpen [0]{}%
\providecommand \bibitemStop [0]{}%
\providecommand \bibitemNoStop [0]{.\EOS\space}%
\providecommand \EOS [0]{\spacefactor3000\relax}%
\providecommand \BibitemShut  [1]{\csname bibitem#1\endcsname}%
\let\auto@bib@innerbib\@empty
\bibitem [{\citenamefont {Preskill}(2013)}]{PreskillSufficient}%
  \BibitemOpen
  \bibfield  {author} {\bibinfo {author} {\bibfnamefont {J.}~\bibnamefont
  {Preskill}},\ }\href@noop {} {\bibfield  {journal} {\bibinfo  {journal}
  {Quantum Inf. Comput.}\ }\textbf {\bibinfo {volume} {13}},\ \bibinfo {pages}
  {181} (\bibinfo {year} {2013})}\BibitemShut {NoStop}%
\bibitem [{\citenamefont {Breuer}\ and\ \citenamefont
  {Petruccione}(2002)}]{Breuer:book}%
  \BibitemOpen
  \bibfield  {author} {\bibinfo {author} {\bibfnamefont {H.-P.}\ \bibnamefont
  {Breuer}}\ and\ \bibinfo {author} {\bibfnamefont {F.}~\bibnamefont
  {Petruccione}},\ }\href@noop {} {\emph {\bibinfo {title} {The Theory of Open
  Quantum Systems}}}\ (\bibinfo  {publisher} {Oxford University Press},\
  \bibinfo {address} {Oxford},\ \bibinfo {year} {2002})\BibitemShut {NoStop}%
\bibitem [{Old()}]{OlderRefs}%
  \BibitemOpen
  \href@noop {} {}\bibinfo {note} {R. J. Schoelkopf and A. A. Clerk and S. M.
  Girvin and K. W. Lehnert and M. H. Devoret, in {\em Quantum Noise in
  Mesoscopic Physics}, NATO Science Series {\bf 97}, 173 (2003); L. Faoro and
  L. Viola, Phys. Rev. Lett. {\bf 92}, 117905 (2004); T. Yuge, S. Sasaki, and
  Y. Hirayama, {\em ibid.} {\bf 107}, 170504 (2011); K. C. Young and K. B.
  Whaley, Phys. Rev. A {\bf 86}, 012314 (2012).}\BibitemShut {Stop}%
\bibitem [{Mik()}]{MikeFF}%
  \BibitemOpen
  \href@noop {} {}\bibinfo {note} {T. J. Green, H. Uys, and M. J. Biercuk,
  Phys. Rev. Lett. {\bf 109}, 020501 (2012); A. Soare, H. Ball, D. Hayes, J.
  Sastrawan, M. C. Jarratt, J. J. McLoughlin, X. Zhen, T. J. Green, and M. J.
  Biercuk, Nature Phys. {\bf 10}, 825 (2014).}\BibitemShut {Stop}%
\bibitem [{\citenamefont {Paz-Silva}\ and\ \citenamefont
  {Viola}(2014)}]{Paz2014}%
  \BibitemOpen
  \bibfield  {author} {\bibinfo {author} {\bibfnamefont {G.~A.}\ \bibnamefont
  {Paz-Silva}}\ and\ \bibinfo {author} {\bibfnamefont {L.}~\bibnamefont
  {Viola}},\ }\href@noop {} {\bibfield  {journal} {\bibinfo  {journal} {Phys.
  Rev. Lett.}\ }\textbf {\bibinfo {volume} {113}},\ \bibinfo {pages} {250501}
  (\bibinfo {year} {2014})}\BibitemShut {NoStop}%
\bibitem [{\citenamefont {\'Alvarez}\ and\ \citenamefont
  {Suter}(2011)}]{Alvarez2011}%
  \BibitemOpen
  \bibfield  {author} {\bibinfo {author} {\bibfnamefont {G.~A.}\ \bibnamefont
  {\'Alvarez}}\ and\ \bibinfo {author} {\bibfnamefont {D.}~\bibnamefont
  {Suter}},\ }\href@noop {} {\bibfield  {journal} {\bibinfo  {journal} {Phys.
  Rev. Lett.}\ }\textbf {\bibinfo {volume} {107}},\ \bibinfo {pages} {230501}
  (\bibinfo {year} {2011})}\BibitemShut {NoStop}%
\bibitem [{Oli()}]{Oliver}%
  \BibitemOpen
  \href@noop {} {}\bibinfo {note} {J. Bylander, S. Gustavsson, F. Yan, F.
  Yoshihara, K. Harrabi, G. Fitch, D. G. Cory, Y. Nakamura, J.-S. Tsai, and W.
  D. Oliver, Nature Phys. {\bf 7}, 565 (2011); F. Yan, S. Gustavsson, J.
  Bylander, X. Jin, F. Yoshihara, D. G. Cory, Y. Nakamura, T. P. Orlando, and
  W. D. Oliver, Nature Commun. {\bf 4}, 2337 (2013); F. Yoshihara, Y. Nakamura,
  F. Yan, S. Gustavsson, J. Bylander, W.D. Oliver, and J.-S. Tsai Phys. Rev. B
  {\bf 89}, 020503 (2014).}\BibitemShut {Stop}%
\bibitem [{Spi()}]{SpinQubits}%
  \BibitemOpen
  \href@noop {} {}\bibinfo {note} {O. Dial {\em et al.}, Phys. Rev. Lett. {\bf
  110}, 146804 (2013); J. T. Muhonen, J. P. Dehollain, A. Laucht, F. E. Hudson,
  T. Sekiguchi, K. M. Itoh, D. N. Jamieson, J. C. McCallum, A. S. Dzurak, and
  A. Morello, Nature Nanotech. {\bf 9}, 986 (2014).}\BibitemShut {Stop}%
\bibitem [{NVs()}]{NVs}%
  \BibitemOpen
  \href@noop {} {}\bibinfo {note} {C. A. Meriles, L. Jiang, G. Goldstein, J. S.
  Hodges, J. Maze, M. D. Lukin, and P. Cappellaro, J. Chem. Phys. {\bf 133},
  124105 (2010); Y. Romach, C. Muller, T. Unden, L. J. Rogers, T. Isoda, K. M.
  Itoh, M. Markham, A. Stacey, J. Meijer, S. Pezzagna, B. Naydenov, L. P.
  McGuinness, N. Bar-Gill, and F. Jelezko, Phys. Rev. Lett. {\bf 114}, 017601
  (2015).}\BibitemShut {Stop}%
\bibitem [{\citenamefont {Paladino}\ \emph {et~al.}(2014)\citenamefont
  {Paladino}, \citenamefont {Galperin}, \citenamefont {Falci},\ and\
  \citenamefont {Altshuler}}]{Paladino}%
  \BibitemOpen
  \bibfield  {author} {\bibinfo {author} {\bibfnamefont {E.}~\bibnamefont
  {Paladino}}, \bibinfo {author} {\bibfnamefont {Y.~M.}\ \bibnamefont
  {Galperin}}, \bibinfo {author} {\bibfnamefont {G.}~\bibnamefont {Falci}}, \
  and\ \bibinfo {author} {\bibfnamefont {B.~L.}\ \bibnamefont {Altshuler}},\
  }\href@noop {} {\bibfield  {journal} {\bibinfo  {journal} {Rev. Mod. Phys.}\
  }\textbf {\bibinfo {volume} {86}},\ \bibinfo {pages} {361} (\bibinfo {year}
  {2014})}\BibitemShut {NoStop}%
\bibitem [{qua()}]{quadratic}%
  \BibitemOpen
  \href@noop {} {}\bibinfo {note} {C. Uchiyama and M. Aihara, Phys. Rev. A {\bf
  66}, 032313 (2002); M. F. Maghrebi, M. Kr\"{u}ger, and M. Kardar,
  arXiv:1508.00582.}\BibitemShut {Stop}%
\bibitem [{\citenamefont {Guttorp}\ and\ \citenamefont
  {Brillinger}(2012)}]{Brillinger}%
  \BibitemOpen
  \bibfield  {author} {\bibinfo {author} {\bibfnamefont {P.}~\bibnamefont
  {Guttorp}}\ and\ \bibinfo {author} {\bibfnamefont {D.}~\bibnamefont
  {Brillinger}},\ }in\ \href@noop {} {\emph {\bibinfo {booktitle} {Selected
  Works of David Brillinger}}}\ (\bibinfo  {publisher} {Springer New York},\
  \bibinfo {year} {2012})\ p.\ \bibinfo {pages} {149}\BibitemShut {NoStop}%
\bibitem [{\citenamefont {Chandran}\ and\ \citenamefont
  {Elgar}(1994)}]{Chandran1994}%
  \BibitemOpen
  \bibfield  {author} {\bibinfo {author} {\bibfnamefont {V.}~\bibnamefont
  {Chandran}}\ and\ \bibinfo {author} {\bibfnamefont {S.}~\bibnamefont
  {Elgar}},\ }\href@noop {} {\bibfield  {journal} {\bibinfo  {journal} {IEEE
  Trans. Signal Proc.}\ }\textbf {\bibinfo {volume} {42}},\ \bibinfo {pages}
  {229} (\bibinfo {year} {1994})}\BibitemShut {NoStop}%
\bibitem [{Not()}]{Note1}%
  \BibitemOpen
  \href@noop {} {}\bibinfo {note} {This inverse problem can be solved by a
  variety of numerical methods. To assess the performance of the protocol with
  minimal computational resources, the simulations presented here utilize
  direct matrix inversion.}\BibitemShut {Stop}%
\bibitem [{Sup()}]{Supplement}%
  \BibitemOpen
  \href@noop {} {}\bibinfo {note} {See Supplemental Material at
  http://link.aps.org/xxx for additional technical details.}\BibitemShut
  {Stop}%
\bibitem [{\citenamefont {Cywi\'{n}ski}(2014)}]{CywinskiOptimal}%
  \BibitemOpen
  \bibfield  {author} {\bibinfo {author} {\bibfnamefont {L.}~\bibnamefont
  {Cywi\'{n}ski}},\ }\href@noop {} {\bibfield  {journal} {\bibinfo  {journal}
  {Phys. Rev. A}\ }\textbf {\bibinfo {volume} {90}},\ \bibinfo {pages} {042307}
  (\bibinfo {year} {2014})}\BibitemShut {NoStop}%
\end{thebibliography}
\end{document}